# The many faces of mobility: Using bibliometric data to measure the movement of scientists

**Nicolás Robinson-Garcia[1,2]*, Cassidy R. Sugimoto[3,4], Dakota Murray[3], Alfredo Yegros-Yegros[4], Vincent Larivière[4,5,6] and Rodrigo Costas[4,7]**

[1] INGENIO (CSIC-UPV)
Universitat Politècnica de València, Spain

[2] School of Public Policy
Georgia Institute of Technology, United States
elrobinster@gmail.com

[3] School of Informatics, Computing, and Engineering
Indiana University Bloomington, USA
sugimoto@indiana.edu; dakmurra@iu.edu

[4] Centre for Science and Technology Studies (CWTS)
Leiden University, The Netherlands
rcostas@cwts.leidenuniv.nl; a.yegros@cwts.leidenuniv.nl

[5] École de bibliothéconomie et des sciences de l'information
Université de Montréal, Canada
vincent.Larivière@umontreal.ca

[6] Observatoire des sciences et des technologies (OST)
Centre interuniversitaire de recherche sur la science et la technologie (CIRST)
Université du Québec à Montréal, Canada

[7] DST-NRF Centre of Excellence in Scientometrics and Science, Technology and Innovation Policy,
Stellenbosch University, South Africa

*Correspondence to: elrobinster@gmail.com

## Abstract

This paper presents a methodological framework for developing scientific mobility indicators based on bibliometric data. We identify nearly 16 million individual authors from publications covered in the Web of Science for the 2008-2015 period. Based on the information provided across individuals' publication records, we propose a general classification for analyzing scientific mobility using institutional affiliation changes. We distinguish between migrants--authors who have ruptures with their country of origin--and travelers--authors who gain additional affiliations while maintaining affiliation with their country of origin. We find that 3.7% of researchers who have published at least one paper over the period are mobile.

Travelers represent 72.7% of all mobile scholars, but migrants have higher scientific impact. We apply this classification at the country level, expanding the classification to incorporate the directionality of scientists' mobility (i.e., incoming and outgoing). We provide a brief analysis to highlight the utility of the proposed taxonomy to study scholarly mobility and discuss the implications for science policy.

# 1. Introduction

Scientific mobility plays an important role in knowledge diffusion and exchange, and is therefore a critical concern for policy makers (OECD, 2010). The notion of mobility is closely associated with the concept of internationalization and is often perceived as beneficial to the scientific enterprise (Wagner & Jonkers, 2017). It is generally related to the phenomenon of migration, with mobility conceived of as a mechanism for global knowledge allocation (Cañibano, 2017) in which some countries *gain* scientific human capital while others *lose* it. This brain drain/gain perspective conceives mobility as a phenomenon of national disparities (Cañibano & Woolley, 2015), where elite countries tend to attract talent to the expense of peripheral countries (Scott, 2015). It is therefore common to find studies on scientific diasporas and brain drain (Davenport, 2004); the influence of immigration in national systems (Levin & Stephen, 1999); and scientists' reasons for immigrating (Stephan, Franzoni, & Scellato, 2016; Ioannidis, 2004), emigrating (Ackers, 2008), or returning to their home country (Jonkers & Tijssen, 2008).

An alternative perspective considers mobility as '[a]n expanded version of the network approach to the migration of highly skilled individuals' (Meyer, 2001, p. 105), in which mobile scientists serve as a bridge between countries. International mobility and collaboration are seen as two facets of the same phenomenon—i.e., internationalization—with the former serving as a trigger for the latter (Kato & Ando, 2017). In this perspective, mobility is a mechanism through which internationalization occurs by enabling scientists' access to transnational collaboration networks. Therefore, while origin countries of mobile scientists may lose human capital, they gain ties with other countries which could, in the long term, be beneficial to their research (Wagner & Jonkers, 2017). In this context, policies promoting short stays abroad and 'international experience' gain significance (Ackers, 2008), as they ensure scientists' return while enabling collaborative linkages with other countries (Cañibano, Fox, & Otamendi, 2016). Consequently, this perspective argues that mobility indicators should reflect both the complexity and the often mutually beneficial aspects of scientific mobility (Cañibano & Woolley, 2015).

Mobility indicators have long been an important component of policy portfolios (Auriol, 2006; Moguérou, di Pietrogiacomo, Da Costa, & Laget, 2006; OECD, 2008). However, these indicators have historically relied on survey data (Akerblom, 1999) and have suffered from issues of data sparsity, consistency, and interoperability (Akerblom, 2000). Despite the importance of scientific mobility for research policy, and the presence of institution and country-level information on scholarly papers, the phenomenon of scientific mobility has been understudied through the lens of bibliometrics. The use of publication data to trace scientists' movements was introduced around 2003 (Laudel, 2003). Subsequent bibliometric studies of mobility focused primarily on the relationship between mobility and indicators of collaboration (Furukawa, Shirakawa, & Okuwada, 2011), production and impact (Aksnes, Rørstad, Piro, & Sivertsen, 2013; Halevi, Moed, & Bar-Ilan, 2015, 2016; Hunter, Oswald, & Charlton, 2009). These studies were often focused on a single journal (Furukawa et al., 2011) or country (Aksnes et al., 2013), a few disciplines (Halevi et al., 2015, 2016), or on a subset of elite scientists (Hunter et al., 2009).

It was not until the development of automatic author name disambiguation techniques that bibliometric approaches were capable of tracking scientific mobility at a large scale (Moed, Aisati, & Plume, 2013; Moed & Halevi, 2014; Robinson-Garcia, Cañibano, Woolley, & Costas, 2016; Sugimoto et al., 2017; Sugimoto, Robinson-Garcia, & Costas, 2016). There is now a sizeable portion of scholarly papers indexed by major bibliometric databases (e.g., Web of Science and Scopus) that include linkages between authors and their specific affiliations. This allows researchers to not only establish institutional ties, but also to discern between authors who have a single affiliation and those that have multiple institutional affiliations. This linkage between authors and affiliations has improved the reliability of author name disambiguation algorithms (Caron & van Eck, 2014), which have been implemented in Scopus since 2011 (Moed et al., 2013), as well as in the in-house version of the Web of Science of CWTS (Caron & van Eck, 2014). These disambiguated datasets have allowed for analyses of mobility flows at the meso- and macro-levels (Sugimoto et al., 2016).

In a recent study (Sugimoto et al., 2017), we analyzed, at the world level, different types of mobile researchers based on the affiliations found on their publications. Here we will further expand on the methodological and conceptual challenges of creating mobility indicators based on bibliometric data. We acknowledge that not all forms of mobility can be captured by bibliometrics. Temporary or short-term mobility, for instance, is not necessarily visible unless associated with a publication. However, several shortcomings of current mobility indicators can be overcome using this method. For example, most mobility studies based on bibliometric indicators tend to follow a brain drain/gain perspective of mobility. This is reflected in the terminology employed in extant indicators, referring to the capacity of countries to ‘attract’ talented scholars (Moed et al., 2013), indistinctive use of the terms ‘migration’ and ‘mobility’ (Moed & Halevi, 2014), and the measurement of ‘returned’ scholars (Robinson-Garcia et al., 2016). Such operationalization provides a relatively limited and reductionist view on the phenomenon of mobility, ignoring the “fluid qualities of globalisation” (Scott, 2015). Moreover, the brain drain/gain perspective ignores the many peculiarities of bibliometric data that suggest many co-existent types of mobility, instead of a homogeneous phenomenon. This study aims to contribute to a better understanding of scientific mobility by describing a new comprehensive taxonomy for mobility using bibliometric data.

# 2. Methods

## 2.1 Data

We used Web of Science Core Collection (SCI, SSCI and A&HCI) to identify global mobility patterns. As has been demonstrated with Scopus data (Moed et al., 2013), links between authors and their affiliations found in bibliometric databases can be used to trace mobility events. Web of Science began linking authors and their affiliations in 2008. Therefore, we retrieved the publication files of all researchers with at least one publication during the 2008-2015 period. All publications for which the link between authors and affiliations has been recorded in the database were included in the analysis regardless of the document types; however, bibliometric indicators (e.g., number of publications) are only calculated for articles and reviews. This study examines nearly 16 million disambiguated individuals who have contributed to more than 13 million distinct publications. Combining all author, publication year, and author-country linkages, the dataset consists of nearly 30 million observations.

## 2.2 Author disambiguation algorithm

In order to study mobility, it is necessary to delineate the entire publication record of each scholar across time. In bibliometric terms, this means using a dataset with disambiguated authors which overcomes the classical issues of homonymy and author name variants that challenge individual-level bibliometrics (Costas, van Leeuwen, & Bordons, 2010; Smalheiser & Torvik, 2009). We employed the author name disambiguation algorithm developed by Caron and van Eck (2014). This algorithm clusters publications using rule-based scorings relying on four categories of the bibliographic data provided in each record:

- *Author*: weighted fields include e-mail address, first and last name, and affiliations (the latter only for publications since 2008);
- *Article*: fields include shared co-authors, funding data, and affiliations (the latter only for publications since 2008);
- *Publication*: fields include subject category and journal; and
- *Citations*: fields include self-citations, co-citations and bibliographic coupling.

Given that the affiliation field is used by the algorithm to identify publications, we can assume an underestimation of mobility, as changes in affiliation lower the probability of clustering two publications for the same researcher together. Limitations in the algorithm may also lead to inaccuracies in clustering East-Asian names and disciplines with hyperauthorship (e.g., High Energy Physics); for these groups, the algorithm tends to split authors, resulting in lower probable mobility rates. The algorithm was initially evaluated using two sets of Dutch researchers: one focused on the author and the other on the aggregate level. To assess the accuracy of the algorithm for the purposes of measuring mobility, we conducted a validation (Sugimoto et al., 2017) using the 2016 version of the Open Researchers and Contributor ID (ORCID) public data file. ORCID provides a dataset of researchers with their educational and occupational affiliations. Therefore, there should be a higher rate of mobility for these scholars than identified through bibliometric means. Our matching exercise found that our bibliometric measure was able to identify a majority of mobile scholars (Sugimoto et al., 2017).

The use of the disambiguation algorithm also allows us to derive two other elements about individual researchers: a) *academic age*, and b) *country of origin*. The year of first publication is used to define academic age (Nane, Larivière, & Costas, 2017), and the author's affiliation on their first publication is used to define their academic origin. This origin is not assumed to be the country of the researcher's birth or nationality, but rather the country in which they likely received their academic training (Robinson-Garcia et al., 2016; Sugimoto et al., 2017).

## 2.3 The many faces of mobility

### Affiliation instances

As a proxy for mobility, we use instances of changes in or multiplicity of affiliations for a single scholar within or across publications. We focus here on international mobility, and define internationally mobile researcher as one who has affiliations with more than one country. We identify three basic types of affiliation instances (i.e., of author-country linkages) for a researcher in a given year:

1) *Single-affiliation instance*: the researcher is affiliated with only one country in all publications in a given year.
2) *Co-affiliation instance*: the researcher is affiliated with two or more countries within a single publication.
3) *Multiple-affiliation instance*: the researcher is affiliated with at least two countries in at least two different publications (e.g., it would typically correspond to two papers with the researcher affiliated with a different country in each paper) published in the same given year.

Permutations of these affiliation instances are what defines the overall *mobility trajectory* of a researcher. Of these, both multiple-affiliation and co-affiliation instances serve to mark international mobility. These instances, however, are not mutually exclusive: researchers may present different affiliation instances during their careers. For example, a researcher may be affiliated with a single country for several years and then establish co- or multiple-affiliations later in the career. Therefore, the mobility trajectory of a researcher is constantly evolving over time based on the chronological combination of affiliation instances. Table 1 presents examples of mobility trajectories for four researchers, all of whom are initially affiliated to a single country and have co- or multiple-affiliations afterwards.

Table 1. Examples of mobility trajectories for four different researchers. Country name in italic denotes country of first publication. (* denotes co-affiliation)

| **RESEARCHER A** | | | **RESEARCHER B** | | |
|---|---|---|---|---|---|
| **Year** | **Country** | **Pubs** | **Year** | **Country** | **Pubs** |
| 0 | *Greece* | 2 | 0 | *Spain* | 1 |
| 1 | Greece | 1 | 1 | Spain | 3 |
| 2 | Belgium | 1 | 2 | Spain | 1 |
| 3 | Belgium | 1 | 3 | United Kingdom | 2 |
| 3 | Greece | 2 | 4 | United Kingdom | 1 |
| 4 | Greece | 1 | 5 | UK*/Italy* | 1 |
| 4 | Belgium | 1 | 6 | Italy | 1 |
| 5 | United Kingdom | 1 | 7 | Italy | 3 |
| 6 | United Kingdom | 3 | | | |
| **RESEARCHER C** | | | **RESEARCHER D** | | |
| **Year** | **Country** | **Pubs** | **Year** | **Country** | **Pubs** |
| 0 | *Spain* | 15 | 0 | *United Kingdom (UK)* | 3 |
| 1 | Spain | 13 | 1 | United Kingdom | 2 |
| 2 | Germany*/Spain* | 1 | 1 | UK*/USA* | 1 |
| 2 | Spain | 14 | 2 | United Kingdom | 3 |
| 2 | USA*/Spain* | 2 | 2 | UK*/USA*/Spain* | 1 |
| 3 | USA*/Spain* | 7 | 3 | UK*/Spain* | 3 |
| 3 | USA | 5 | 4 | UK*/Spain* | 2 |
| 4 | USA*/Spain* | 11 | 5 | UK*/Spain* | 3 |
| 4 | USA | 9 | 6 | UK*/Spain*/France* | 2 |

In Table 1, researchers A and B are examples of researchers who have migrated to different countries (other than their country of origin). Researcher A has ruptures with her country of origin (Greece) at different points in time (i.e., years 2 and 5) during which she is affiliated with two new countries (i.e., Belgium and the UK). Research B instead ruptures with their origin country (Spain), affiliates within the U.K., and briefly holds a co-affiliation with the U.K. and Italy until affiliating solely within Italy. In the case of researcher A, an instance of multiple affiliation is identified between years 2 and 3, while researcher B has a co-affiliation instance at year 5 (probably a transitional phase between the United Kingdom and Italy). In the final years of the analysis, researcher B has lost ties to both their country of origin (Spain) and their country of second affiliation (UK). Researcher C shows indications of co-affiliation in some papers but not in others, while researcher D always presents co-affiliation instances. Neither of these two researchers loses ties with their country of origin, although both exhibit linkages with different countries over the time (e.g., researcher C has some papers within a given year without the country of origin, while researcher D retains ties to their country of origin on all publications). These examples demonstrate an important element in operationalizing mobility through bibliometric data: the predominant role of the publication *year*. Thus, the mobility trajectory of researcher is determined by the combination of affiliation instances across years, and not simply across publications. Thus, if a researcher migrates but does not publish in the following year, then they cannot be classified as a migrant; we discuss this more thoroughly later.

Table 1 also suggests that *directional* notions of migration (largely associated with brain drain/gain metaphors) may not fully capture all the mobility trajectories of researchers. For example, while researchers A and B seem to have moved more permanently from one country to another, researchers C and D constantly retain affiliations to their countries of origin. Therefore, we cannot assume that these researchers have migrated (i.e., breaking ties with their country of origin and go to another country) but rather that they are engaging in short-term research visits, part-time appointments, temporary stays, or remote employment. These types of short-term mobility characterize a large volume of mobility trajectories (Børing, Flanagan, Gagliardi, Kaloudis, & Karakasidou, 2015; Cañibano, Otamendi, & Solís, 2011; Markova, Shmatko, & Katchanov, 2016). Notions such as *return* would be applicable to researchers A and B (e.g., A returned to Greece in year 3, although left again in year 5; while researcher B never returned) and brain/gain indicators could be compiled (e.g., Greece lost researcher A to Belgium and the United Kingdom; while Spain and the United Kingdom lost researcher B to Italy). However, these notions would not apply to researchers C and D as we cannot discern whether they have fully departed from their country of origin at any point in time.

## Mobility events

In practical terms, the mobility trajectory of a researcher can be determined by different *mobility events.* A *mobility event* refers to each of the different possible permutations of affiliation instances that a researcher can have between two points in time. We propose the following basic notation to identify different mobility events:

- $C_1$: first[1] country of affiliation of an author. $C_2$: any other country (or set of countries) of affiliation of an author which is different from $C_1$.
- $t_n$ and $t_{n+1}$: refer to a particular point in time (in this case, the year), in which the links of an author to a country (or set of countries) has been tracked. Specifically, at $t_0$ the affiliated country(ies) is(are) considered the *country(ies) of origin* of the author[2]. In addition, $t_{n+1}$ refers to the following tracked point in time (year) in the publication record of the researcher.
- $C_1$; $C_2$: the author has at least one publication affiliated with $C_1$ and another publication affiliated with $C_2$ in the same $t_n$.
- $C_1$*; $C_2$*: the author has at least one publication co-affiliated with $C_1$ and $C_2$ at the same $t_n$.

There are also two particular elements that can be related to mobility events over time. These elements will further refine the understanding of scholars' mobility trajectories.

- *Directionality*: indicates whether it is possible to reliably establish whether the author has been chronologically affiliated first to $C_1$ and then to $C_2$. We rely on the publication year (due to publication delays) as an imperfect temporal unit of analysis with the lowest level of granularity in Web of Science to track chronological changes.
- *Country rupture*: when a researcher's country(ies) at $t_n$ are not found among the affiliations of the researcher at $t_{n+1}$. In other words, there is a *rupture* in the countries between $t_n$ and $t_{n+1}$.

In Table 2 we summarize all conceptually possible mobility events that a researcher can have, allowing us to establish a taxonomy of different mobility events. This taxonomy defines up to 15 different mobility events, which are characterized by the affiliation instances of authors between two points in time (denoted as $t_n$ and $t_{n+1}$), although not all event types have information in the second point in time ($t_{n+1}$) (e.g. E1, E6, and E7 would refer to events happening at just one point in time. E6 and E7 would include the cases in which researchers are already affiliated to more than one country in their first year of publication).

Table 2. Taxonomy of mobility events tracked through affiliation instances between $t_n$ and $t_{n+1}$

| **Mobility Event** | **$t_n$** | **$t_{n+1}$** | **Directionality** | **Country rupture** |
|---|---|---|---|---|
| E1 | $C_1$ | | No | No |
| E2 | $C_1$ | $C_1$ | No | No |
| E3 | $C_1$ | $C_2$ | Yes | Yes |
| E4 | $C_1$ | $C_1$; $C_2$ | Yes | No |
| E5 | $C_1$ | $C_1$*; $C_2$* | Yes | No |
| E6 | $C_1$; $C_2$ | | No | No |

[2] A researcher could have two (or more) countries of origin if she is affiliated with more than one country within her first year of publication.

| Mobility Event | $t_n$ | $t_{n+1}$ | Directionality | Country rupture |
|---|---|---|---|---|
| E7 | $C_1$*; $C_2$* | | No | No |
| E8 | $C_1$*; $C_2$* | $C_1$ | No | No |
| E9 | $C_1$*; $C_2$* | $C_2$ | No | No |
| E10 | $C_1$; $C_2$ | $C_1$; $C_2$ | No | No |
| E11 | $C_1$; $C_2$ | $C_1$*; $C_2$* | No | No |
| E12 | $C_1$; $C_2$ | $C_1$ | No | No |
| E13 | $C_1$; $C_2$ | $C_2$ | No | No |
| E14 | $C_1$*; $C_2$* | $C_1$; $C_2$ | No | No |
| E15 | $C_1$*; $C_2$* | $C_1$*; $C_2$* | No | No |

In Table 2, events E1 and E2 do not signal any mobility, as in these cases authors would be affiliated to only one country. In the case of E3, it corresponds to a mobility event in which a researcher is affiliated to two different countries at two different points in time and for which some directionality from one country to another can be established. This event (E3) points to a *migration* in the mobility trajectory of a researcher. Mobility events E4 and E5 signal trajectories in which a researcher starts in a country ($C_1$), later establishes linkages with $C_2$ but without a clear rupture with $C_1$. Finally, mobility events E6 to E15 refer to events in which linkages with more than one country are identified for a given researcher but without clear directionality between countries.

## Mobility classification of researchers

Based on the taxonomy above it is possible to establish different individual-level *mobility classes* based on the presence of specific mobility events in the profile of researchers. We propose a system of classification to define mobility at the level of the individual researcher, as measured across their overall publication output:

- *Not mobile*: researchers lacking any mobility event showing affiliation instances with more than one country (researchers only having mobility events E1 and E2, table 2).
- *Migrants*: researchers with a directional mobility event and a point of rupture with their country of origin (i.e., they have a mobility event E3 at any point in time, table 2). Moreover, from a country point of view, migrants can be further classified as *immigrants* (from the perspective of $C_2$ countries) or *emigrants* (from the perspective of $C_1$ countries).
- *Travelers (directional):* researchers with directionality but no rupture with their country of origin (mobility events E4 and E5, table 2). For countries having this type of mobile scholars, due to the directionality, it is also possible to establish which countries have *outgoing* travelers ($C_1$ countries) and *incoming* travelers ($C_2$ countries).
- *Travelers (non-directional):* researchers with at least one mobility event but no directionality and no rupture with their country of origin (mobility events E6-E15 in table 2). Note that non-directional travelers must have more than one country of origin and consistently be affiliated to at least one of their countries of origin without adding additional countries.

Following the examples set in Table 1 and considering our mobility classification, we would classify researchers A and B as migrants, while C and D are travelers (directional). We acknowledge here that the classification of researchers proposed above is not absolute; based on the taxonomy presented in Table 2, other classes could also be discussed. For example, mobility event E4 could be seen as slightly different from E5, since E4 requires the researcher to have published at least two publications at $t_{n+1}$, one with $C_2$ and another one with $C_1$. This could be seen as a potentially stronger maker of mobility than the one captured by E5 (in which at $t_{n+1}$ the researcher is co-affiliated with both $C_1$ and $C_2$). Similarly, although we see mobility events from E8 to E15 as marking travelers (non-directional), one could argue that in E8, E9, E12 and E13 there is some form of *inverse directionality* since these events involve the loss of at least one country in the mobility trajectory of the researcher. Therefore, there are many different types of mobility that could be derived from Table 2. The strength of the proposed taxonomy is its capacity to reduce the complexity depicted in Table 2 to a smaller and more intuitive set of typologies.

# 3. Results

## 3.1 General Results

A total of 15,931,221 researchers were identified in the 2008-2015 period with at least one publication indexed in the Web of Science Core Collection (SCI, SSCI and A&HCI). The majority of these researchers (96.3%) showed no evidence of international mobility, while 3.7% showed some kind of international mobility. The most common type of mobile researchers identified are *travelers* (*non-directional*) (1.4%) followed closely by *travelers* (*directional)* (1.3%). *Migrants* were the least prevalent type of mobile researchers. Table 3 offers a general overview of number of researchers analyzed, by type of mobility class.

Table 3. Descriptive values of researchers by mobility class for the 2008-2015 period

| | **All** | **Not mobile** | **Mobile** | **Migrants** | **Travelers (directional)** | **Travelers (non-directional)** |
|---|---|---|---|---|---|---|
| **% of total** | 100.0% | 96.3% | 3.7% | 1.0% | 1.3% | 1.4% |
| **% of mobile** | -- | 0.0% | 100.0% | 27.3% | 35.9% | 36.8% |
| **Total** | **15,931,221** | **15,335,327** | **595,894** | **162,519** | **213,810** | **219,565** |

Table 4 shows the distribution of researchers based on the number of publications produced overall and by mobility class. Almost 75% of all identified researchers produced only one or two citable documents (articles or reviews). In the case of non-mobile researchers, this share increases to 77%. Mobile researchers with one or two publications represent 26% of the whole share with almost 40% of them having produced more than 10 papers. The largest share of mobile researchers with more than 10 publications is observed for directional travelers. These represent almost 60% of the total number of travelers. Migrants with more than 10 publications represent 37% of the total number of migrants. In both cases, the share of researchers increases as the number of publications by researcher increases. The only exception observed is for non-directional travelers. While more than 20% of these have more than 10 publications (as opposed to 14% of non-mobiles), 56% of non-directionals have published one or two papers. This indicates that this group of mobile researchers may be affected by limitations of the author name disambiguation algorithm, which tends to split authors when the probability of publications belonging to the same author is low. It is important to note that, while figures on production and number of papers only include citable documents (articles and reviews), we have tracked mobility using all document types. This explains why some migrants (for whom at least two publications in different years are required) exhibit a production of only one document.

When analyzing citation impact indicators, we confirm a phenomenon observed elsewhere (Larivière & Costas, 2016), that is, the increasing number of publications leads to higher citation rates both in terms of Mean Normalized Citation Score (MNCS) and share of highly cited papers (top 10% HCP). In this case, the highest overall rates are observed for migrants, followed by directional travelers, non-directional travelers, and finally, not mobile researchers.

Table 4. Main indicators on mobility at the individual level – by number of publications

| | All researchers | | | Mobile researchers | | | Not mobile | | |
|---|---|---|---|---|---|---|---|---|---|
| **# Pubs** | **# researchers** | **10% HCP** | **MNCS** | **# researchers** | **10% HCP** | **MNCS** | **# researchers** | **10% HCP** | **MNCS** |
| **1** | 8,638,246 | 8.8% | 0.94 | 100,700 | 13.2% | 1.29 | 8,537,546 | 8.8% | 0.94 |
| **2** | 1,358,186 | 8.6% | 0.94 | 47,457 | 10.6% | 1.10 | 1,310,729 | 8.6% | 0.93 |
| **3** | 690,261 | 9.4% | 1.01 | 38,627 | 11.4% | 1.20 | 651,641 | 9.3% | 1.00 |
| **4** | 440,364 | 9.9% | 1.05 | 32,985 | 12.0% | 1.21 | 407,380 | 9.8% | 1.04 |
| **5** | 310,545 | 10.3% | 1.08 | 28,658 | 12.5% | 1.26 | 281,888 | 10.1% | 1.06 |
| **6** | 231,951 | 10.7% | 1.11 | 25,662 | 13.3% | 1.29 | 206,283 | 10.4% | 1.09 |
| **7** | 181,201 | 10.9% | 1.12 | 22,392 | 13.5% | 1.31 | 158,810 | 10.5% | 1.09 |
| **8** | 146,403 | 11.1% | 1.13 | 20,380 | 13.6% | 1.32 | 126,017 | 10.6% | 1.10 |
| **9** | 121,268 | 11.2% | 1.14 | 18,523 | 13.9% | 1.34 | 102,746 | 10.7% | 1.10 |
| **10** | 101,137 | 11.4% | 1.17 | 16,475 | 14.3% | 1.37 | 84,662 | 10.8% | 1.13 |
| **>10** | 1,118,454 | 14.4% | 1.50 | 229,193 | 16.3% | 1.60 | 889,262 | 13.9% | 1.47 |
| | **Mobile researchers by mobility classes** | | | | | | | | |
| | **Migrants** | | | **Travelers (directional)** | | | **Travelers (non-directional)** | | |
| **# Pubs** | **# researchers** | **10% HCP** | **MNCS** | **# researchers** | **10% HCP** | **MNCS** | **# researchers** | **10% HCP** | **MNCS** |
| **1** | 3,047 | 12.1% | 1.16 | 2,566 | 12.5% | 1.24 | 95,087 | 13.2% | 1.30 |
| **2** | 15,713 | 10.7% | 1.15 | 10,575 | 10.5% | 1.07 | 21,169 | 10.5% | 1.08 |
| **3** | 13,952 | 11.7% | 1.22 | 12,088 | 11.4% | 1.20 | 12,587 | 11.1% | 1.17 |
| **4** | 12,801 | 12.4% | 1.25 | 11,575 | 11.7% | 1.18 | 8,609 | 11.6% | 1.18 |
| **5** | 11,742 | 13.2% | 1.31 | 10,413 | 12.5% | 1.28 | 6,503 | 11.5% | 1.15 |
| **6** | 10,751 | 14.2% | 1.31 | 9,585 | 12.6% | 1.28 | 5,326 | 12.4% | 1.27 |
| **7** | 9,470 | 14.4% | 1.34 | 8,533 | 13.1% | 1.30 | 4,389 | 12.4% | 1.26 |
| **8** | 8,716 | 14.6% | 1.37 | 7,974 | 13.0% | 1.30 | 3,690 | 12.4% | 1.22 |
| **9** | 7,872 | 15.2% | 1.44 | 7,452 | 13.1% | 1.29 | 3,199 | 12.5% | 1.22 |
| **10** | 6,898 | 15.7% | 1.47 | 6,758 | 13.2% | 1.30 | 2,819 | 13.5% | 1.28 |
| **>10** | 59,889 | 17.4% | 1.65 | 125,294 | 16.0% | 1.60 | 44,010 | 15.6% | 1.56 |

The dependency on research output is even more evident in Figure 1, where we observe that the population researchers with more publications also consists of a greater share of mobile researchers. As expected, non-mobile researchers tend to exhibit lower production levels—the chance to identify mobility is lower for those with low number of publications. *Non-directional travelers* are characterized by having co-affiliation in most of their affiliation instances. As co-affiliation is less dependent on the number of publications (just one paper in which the author appears linked to several countries is enough to show evidence of co-affiliation), this group exhibits a lower production level. Multiple affiliation instances have a stronger dependence on the number of publications as it requires at least two publications to show a full change of international affiliations. Hence, *migrants* tend to have a higher production level. Finally, researchers with both co-affiliation and multiple affiliations demonstrate the highest level of productivity. This seems is the case for *directional travelers*, who show the highest production levels (although issues such as seniority may also play a role here).

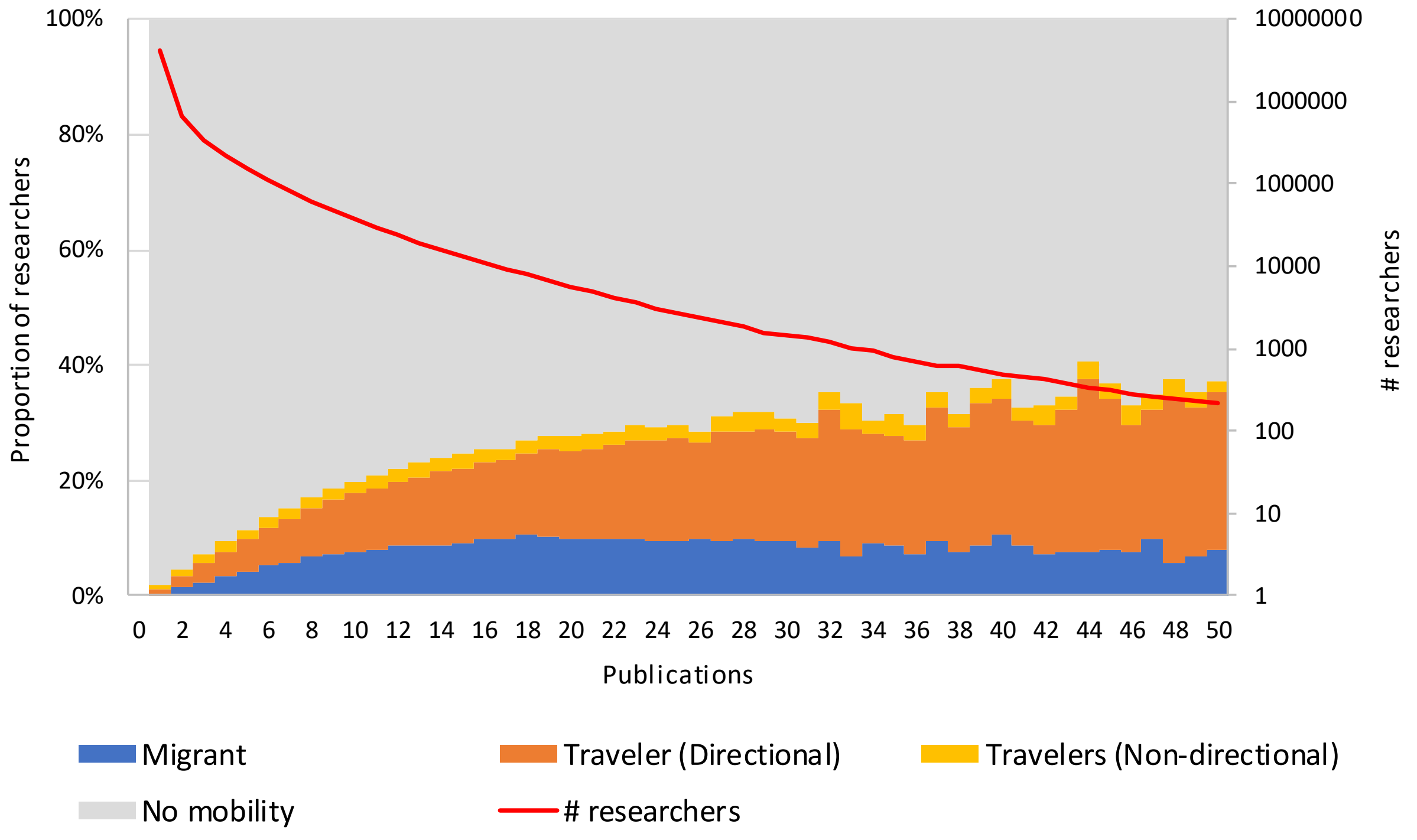

Figure 1. Proportion of researchers by number of publications (left axis) and number of researchers in logarithmic scale by number of publications (right axis)

Citation impact differences seem to be consistent when controlling by number of publications. Figure 2 confirms the pattern observed in Table 4—given a minimum threshold of publications, migrants are consistently the group with highest citation impact. Travelers have similar citation rates, with directional travelers showing slightly higher values than non-directionals. Non-mobile researchers, the largest group, exhibit the lowest MNCS values regardless of their publication output.

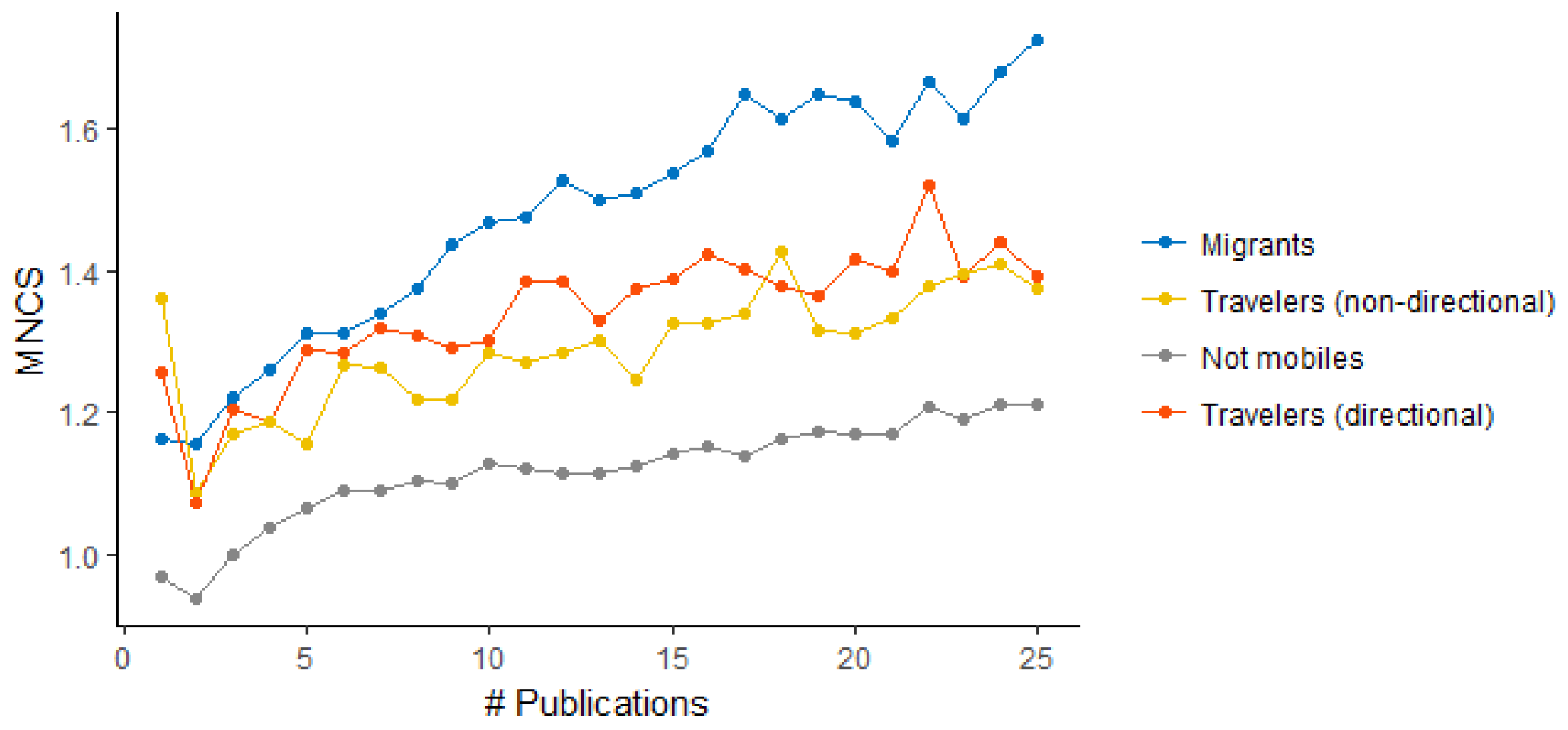

Figure 2. MNCS values at the researcher level by mobility type controlling by number of publications

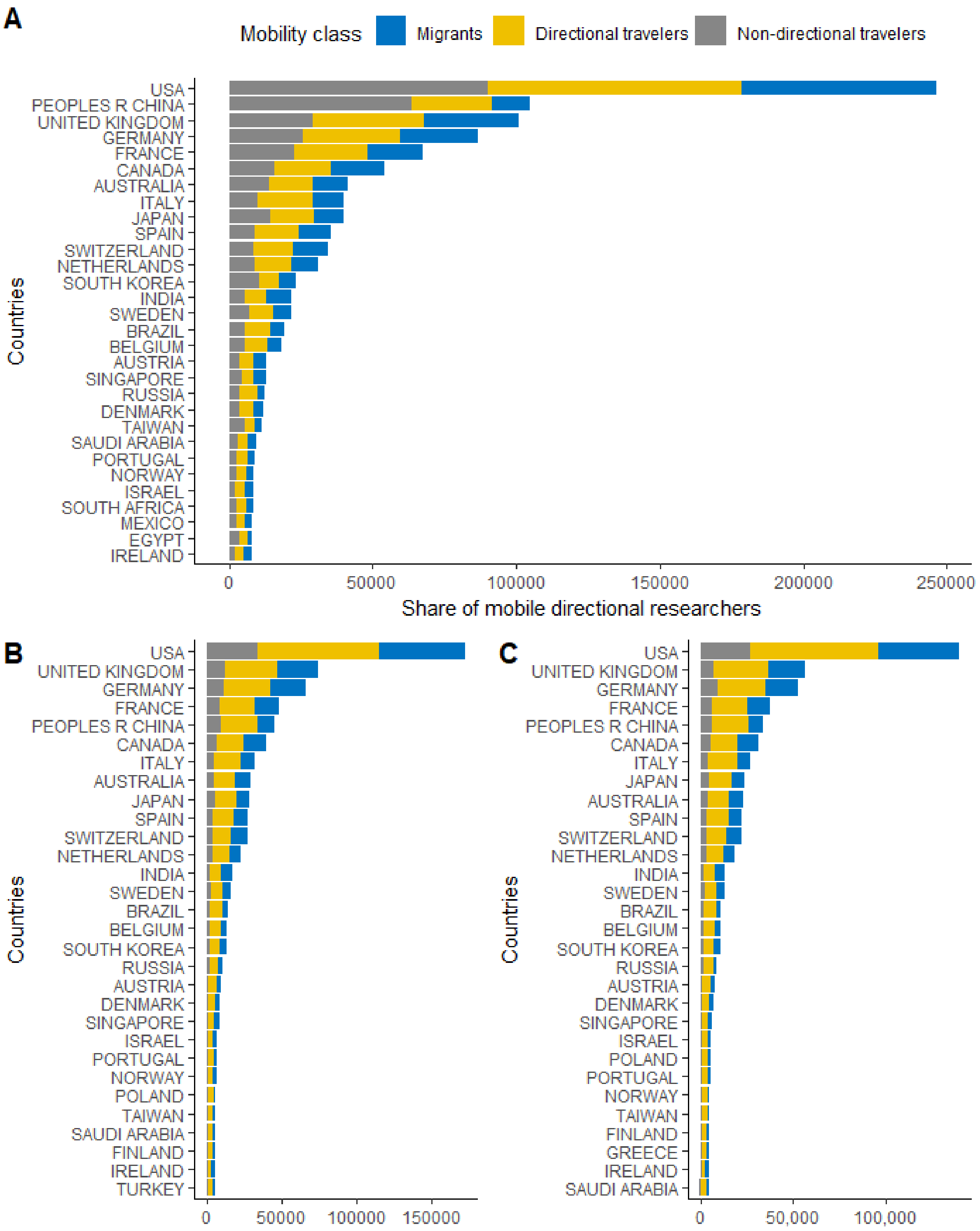

Figure 3. Number of mobile researchers per country broken down by mobility classes. A) All mobile researchers, B) Mobile researchers with ≥5 publications, C) Mobile researchers with ≥10 publications. Only top 30 countries with highest number of mobile researchers are shown

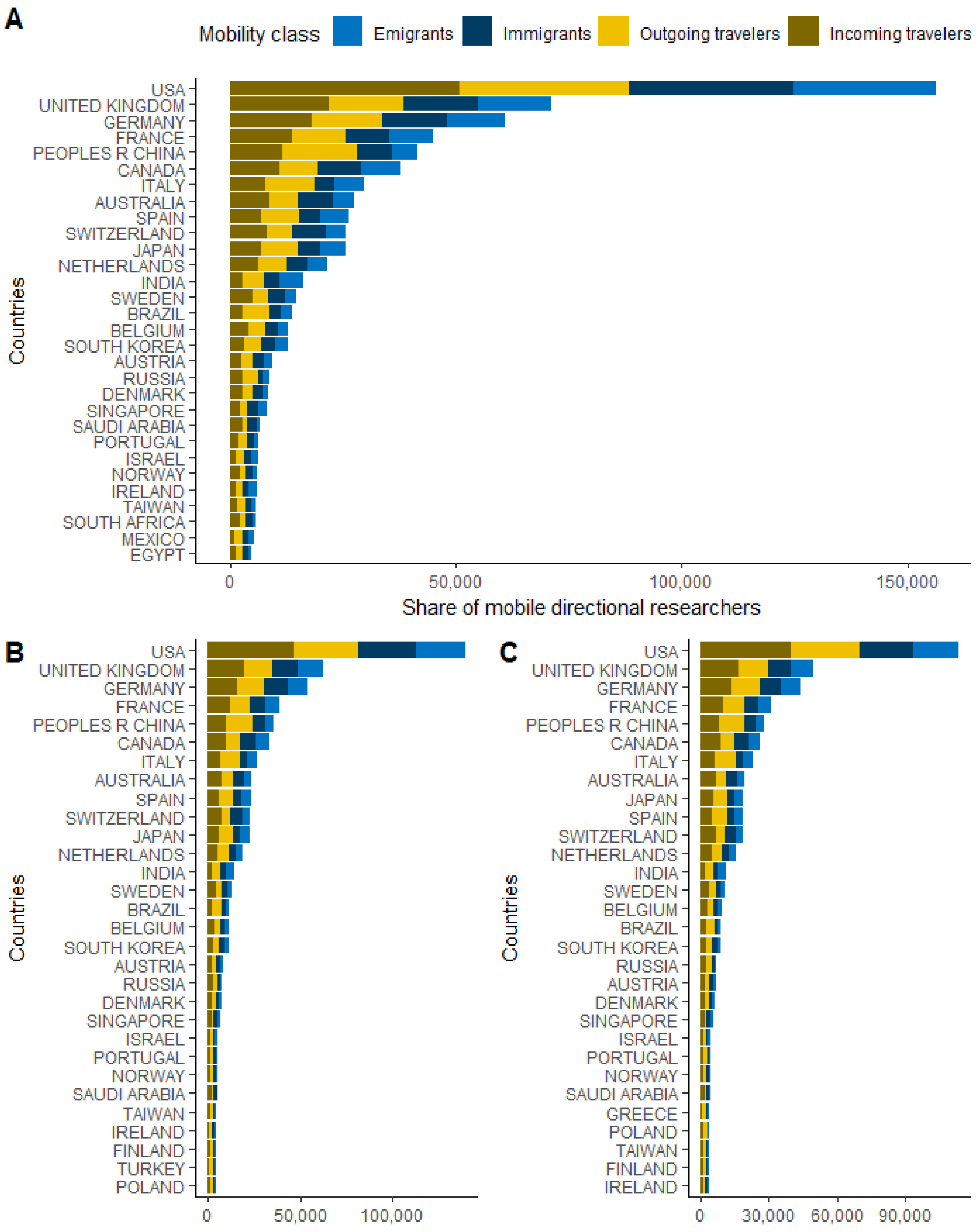

Figure 4. Number of mobile directional researchers per country broken down by mobility classes. A) All mobile researchers, B) Mobile researchers with ≥5 publications, C) Mobile researchers with ≥10 publications. Only top 30 countries with highest number of mobile directional researchers are shown

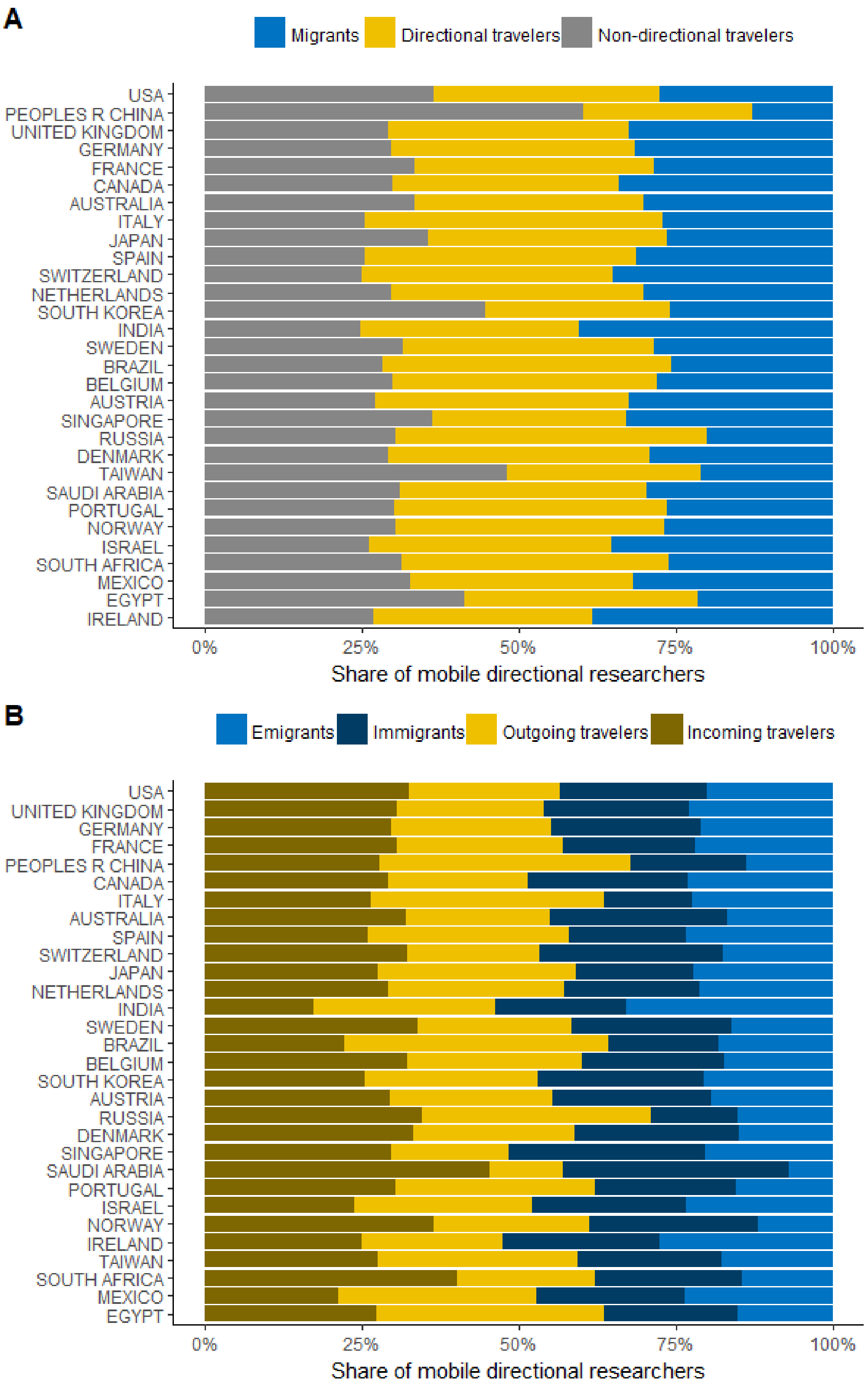

Figure 5. Mobility distribution of scientific workforce for top 30 countries with the largest number of mobile researchers

## 3.2 Analysis by country

In this section we put into practice the taxonomy described above by providing some figures on the distribution of researchers by mobility type at the country level. The purpose of this section is not to offer an in-depth analysis of the phenomenon of international scientific mobility, but instead to illustrate the potential of the proposed taxonomy for informing science policy.

Figure 3 presents the numbers of mobile researchers per country, with a breakdown of the three different mobility classes. In this figure, we include any linkage between countries and researchers. Thus, migrants refers both to researchers that migrated *to* and researchers who migrated *from* a country; the same rule applies to travelers. The USA, China, and the United Kingdom are associated with the largest number of mobile researchers, while countries like Mexico, Egypt, and Ireland have far fewer. This visualization shows only raw figures, and so more populous countries will tend to show larger numbers of all classes of mobile researchers. Figures 3B and 3C include two minimum publications thresholds of at least 5 and 10 publications respectively, to show potential changes derived from mistakenly split authors derived from the author name disambiguation algorithm. As observed, Asian countries are the more likely to be affected by this limitation of the algorithm.

In Figure 4, we show a perspective that emphasizes directionality. We restrict our analysis to typologies of migrants and directional travelers. In this case we distinguish migrant scholars as *emigrants* (those that have left the country at some point) and *immigrants* (those that started in a different country and arrive to the given country at some point). Similarly, we subdivide travelers into *outgoing travelers* (those that started their production in a specific country although later on their established new linkages with other countries) and *incoming travelers* (those travelers that started their output in another country and established a new affiliation linkage with the given country but without breaking ties with their initial country). Non-directional travelers are excluded at the country level because previously-mentioned limitations of the author name disambiguation algorithm restrict any interpretations made at this aggregation level. Again, Figure 4B and 4C include 5 and 10 publications thresholds, respectively. Although some differences with Figure 4A are observed in the list of countries, these are minimized once non-directional travelers are removed.

Generally, the distribution by mobility class is relatively stable by country, but with notable exceptions (Figure 5A). Countries like China, South, Korea, Taiwan, and Egypt have larger shares of non-directional travelers. This effect is minimized when introducing a minimum publication threshold as noted in subsection 3.1. In Figure 5B we remove non-directional travelers and analyze countries' composition, distinguishing between incoming and outgoing researchers. Some cases stand out: e.g., the majority of Saudi Arabia's travelers are incoming (79.8%), but they also have a larger share of immigrants than emigrants (84%). To a lesser extent, a similar pattern is observed for South Africa, with 64.7% of its incoming travelers and 62.2% of its migrants being immigrants. 58.7% of China's travelers are outgoing and 57.4% of its migrants are incoming. On the other hand, we find countries with larger shares of emigrants and outgoing travelers. The most extreme case is India, where 61.3% of its migrants are researchers emigrating from the country and 37.6% of its travelers are connecting with other countries. Ireland, Brazil, Mexico, and Spain also show similar patterns.

Next, we combine these two mobility types (i.e., directional travelers and migrants). Figure 6 shows the difference between emigrants and immigrants as well as between outgoing and incoming travelers. In both cases, a zero value indicates a given country has the same share of outgoing and incoming mobile researchers. The indicator presented in Figure 6A mimics the brain gain/drain model, where a negative value shows a loss of human capital, and a positive value shows an attraction of foreign scholars. The indicator used in Figure 6B does not have a similarly negative connotation—a negative value here simply indicates that travelers from the country of origin are connecting with other countries. These two perspectives consider only on the circulation of human capital and do not consider relative benefits that may be derived from each mobility class (migration and traveling) such as forming collaborative ties and sharing social capital. By considering both emigration/immigration and outgoing/ incoming travelers, a more nuanced perspective of mobility emerges. For example, we observe that although Italy is not the country with the largest share of emigrants, it has a stronger imbalance compared to other countries (e.g., Spain). Another interesting case is that of France, which despite showing figures near to zero for both rates, its emigrating-immigrating difference is negative (-0.02) while its outgoing-incoming difference is positive (0.07). In the case of Singapore, although presenting with positive patterns for both indicators, there are important differences in scale.

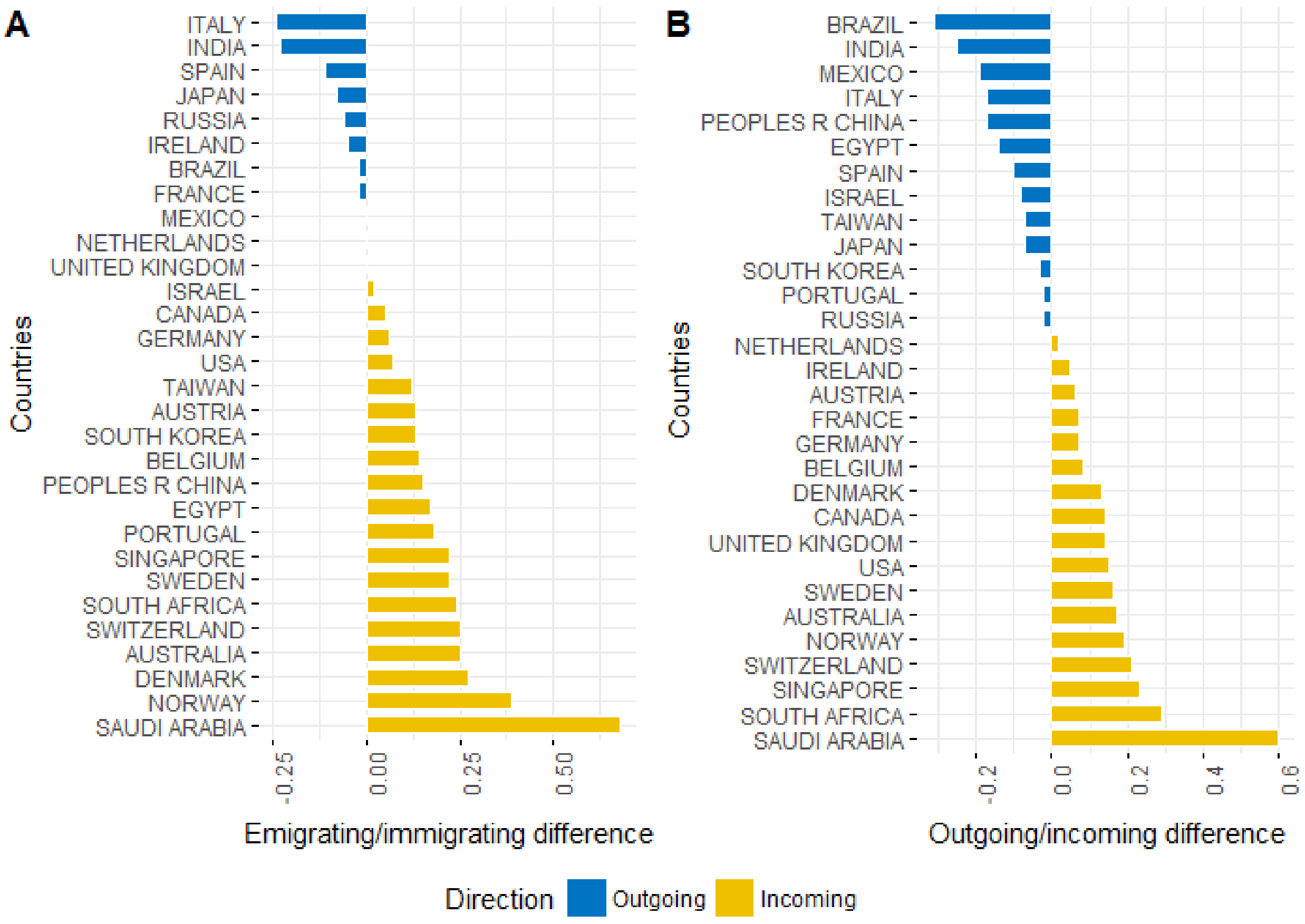

Figure 6. A) Migrants. B) Travelers rate for top 30 countries with largest number of mobile researchers

Figure 7 visualizes both emigrating/immigrating and outgoing/incoming differences, providing a deeper understanding of the relation between these mobility classes at the country-level. As could be expected, these two indicators are highly correlated (R=0.84); that is, countries attracting scientists attract them either by integrating them into their national systems or establishing ties with them. We further confirm some of the patterns already addressed, such as the high attraction of researchers from several Middle East countries such as Saudi Arabia, Qatar, or Iraq. Leading scientific countries such as the USA, Germany, or Switzerland show positive rates of attraction, as they are receiving countries for both traveling and migrant scholars. An interesting case is the United Kingdom, which shows a relatively balanced emigrating-immigrating difference but tends to receive more traveling scholars than it sends. Countries like Iran, India, Ukraine, and Cameroon are all positioned in the lower left quadrant, with more emigrating scholars and outgoing travelers connected with other countries. Southern European countries such as Greece, Italy and Spain, along with most South American countries, have more outgoing researchers than incoming in both mobility classes.

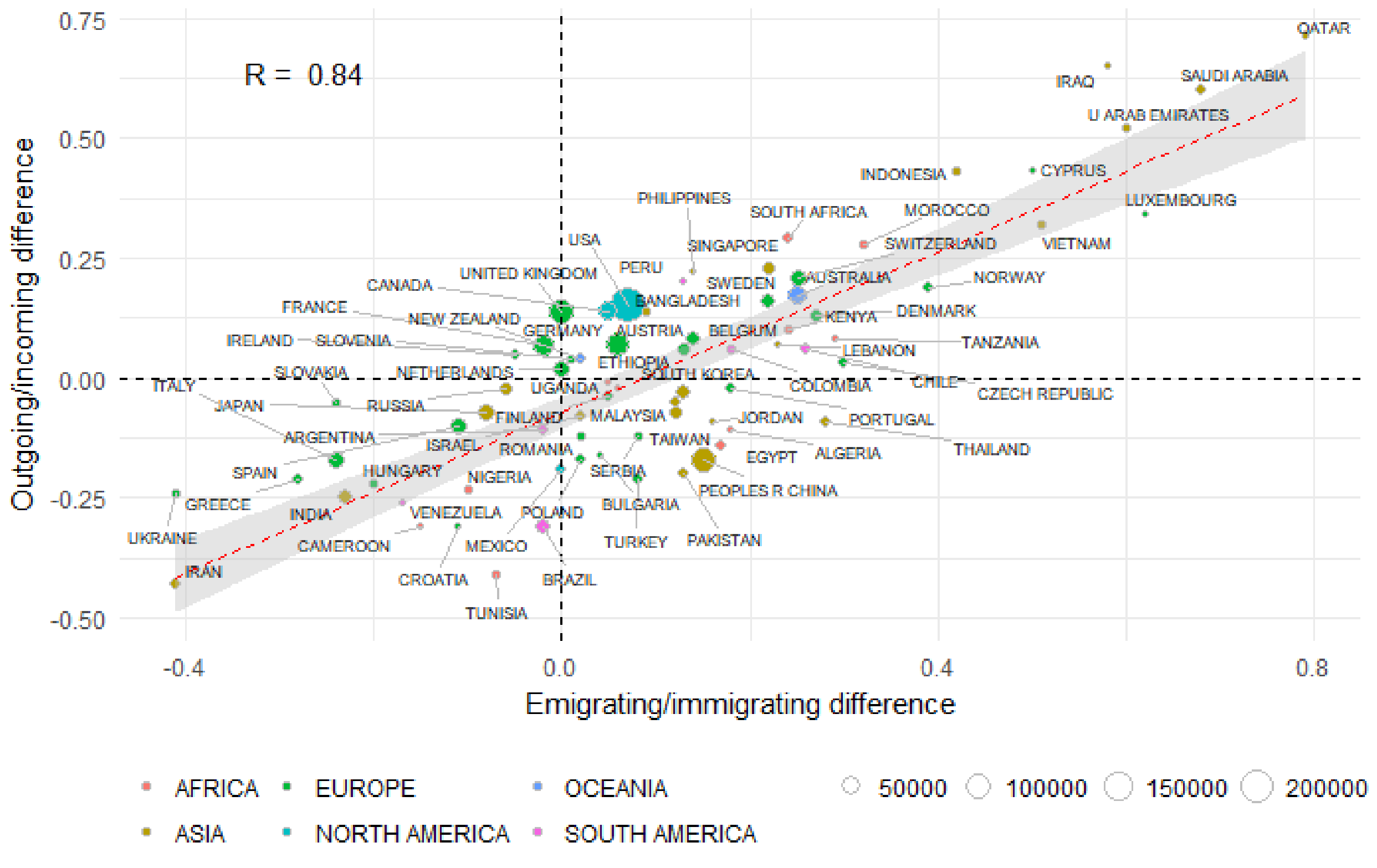

Figure 7. Migrants vs. Travelers. Countries with at least 1000 mobile researchers

Although analyzing the position of each country is out of the scope of this study, it is interesting to note those countries with disparities between migration and circulation rates; these represent special cases that warrant further analysis. For example, China is a country that is gaining migrants and simultaneously having a larger share of outgoing travelers who establish linkages with other countries. On the opposite side we observe the case of France, a country with a larger share of foreign travelers than those originating in the country, but a larger share of emigrants than immigrants.

# 4. Discussion

This paper presents a methodological framework to study international scientific mobility using bibliometric data. This work extends previous research on macro-level mobility indicators (Moed & Halevi, 2014, 2014) using the assumptions of a circulation-based theory of scientific mobility (Cañibano & Woolley, 2015). There are, of course, several limitations to this approach. The main caveat of our bibliometric approach is that we only track mobility that manifests in publications. One result of this limitation is that classification of mobility is dependent on researchers' research output. Given that publications are the signaling device for tracing mobility, it is reasonable to assume that the probability of a mobility event will increase positively with increased production (Abramo, D'Angelo, & Solazzi, 2011). This method is also likely to underrepresent short-term stays or other employment opportunities that do not result in a publication or which may not warrant adding an affiliation. Reliance on low-resolution publication data limits the tracking of mobility to the level of the year, thus obscuring more high-frequency mobility. Delays in publication also means that the observed mobility is delayed from the actual mobility. Furthermore, data is limited to those publications in indexed databases, which underrepresents certain countries, disciplines and languages (Sugimoto & Lariviere, 2018). Related to these concerns are the limitations of the author-name disambiguation algorithm. The algorithm used in this research (Caron & van Eck) uses stability in affiliation as a criterion for disambiguation; therefore, authors with high levels of mobility may be "split" into two or more different identities, resulting in an underestimation of mobility. The limitations of this disambiguation algorithm should be further analyzed, including identifying those countries or fields in which it may present higher inconsistencies. However, comparison with ORCID data performed in Sugimoto et al. (2017) suggest that mobility observed in bibliometric data is similar to that observed in ORCID data, although ORCID had a higher share of mobile researchers. Finally, we take as the country of origin the first country of publication; we may thereby overlook scholars who moved prior to their first publication. In all these cases, we can consider our indicator to be a conservative estimate of mobility, whereby underestimating mobility is more likely than overestimation.

Despite these limitations, our approach allows for much larger-scale mobility analyses than previous studies that relied on survey data, registries, or CVs. Studying scientific mobility from a bibliometric perspective allows for multi-disciplinary, large-scale, global, longitudinal, and contemporary analyses of the exchange of publishing scholars (Sugimoto et al., 2016). The data can be integrated with other bibliometric indicators such as scholarly impact (Sugimoto et al., 2017), collaboration indicators (Chinchilla-Rodríguez et al., 2017), and with sociodemographic data (Azoulay, Ganguli, & Zivin, 2016). Furthermore, the construction of the proposed mobility classification demonstrates the richness of bibliometric data, which favors more nuanced approaches compared with previous and more simplistic analyses. Future studies can take advantage of this richness by introducing network-based approaches to study how countries are related by mobility flows. Our results also provide a foundation for further enquiries into the explanatory factors behind the positioning of countries and their capacity to attract or repel certain populations of researchers (e.g., the enforcement of migration policies, recruitment programs, etc.). Previous understandings of scientific mobility were binary—considering only immigrants and emigrants; our taxonomy is more richly layered, and offers a more comprehensive understanding of the intricacies of mobility. We attempt to craft more nuanced indicators that depict the many faces of mobility. This macro-level prosopographic approach can easily be applied at multiple levels of analysis—demonstrating scientific mobility within nations or localities, across institution and sectors, and between scientific disciplines.

## Acknowledgments

Preliminary results of this paper were presented at the STI Conference 2017 held in Paris and at the I Reunión de Servicios de Evaluación Científica en los Vicerrectorados de Investigación held in Granada. The authors would like to thank Carolina Cañibano and Henk F. Moed for fruitful discussions and comments. We also acknowledge comments made by two anonymous reviewers and the guest editor. Funding from the South African DST-NRF Centre of Excellence in Scientometrics and STI Policy (SciSTIP) is also acknowledged.

# References

Abramo, G., D'Angelo, C. A., & Solazzi, M. (2011). The relationship between scientists' research performance and the degree of internationalization of their research. *Scientometrics*, *86*(3), 629–643. https://doi.org/10.1007/s11192-010-0284-7

Ackers, L. (2008). Internationalisation, mobility and metrics: A new form of indirect discrimination? *Minerva*, *46*(4), 411–435. https://doi.org/10.1007/s11024-008-9110-2

Akerblom, M. (1999). Mobility of highly qualified manpower: A feasibility study on the possibilities to construct internationally comparable indicators. In *Workshop of OECD Focus Group on Human Resources and Mobility. Paris* (Vol. 10).

Akerblom, M. (2000). Constructing Internationally Comparable Indicators on the Mobility of Highly Qualified Workers: A Feasibility Study. *STI-Science Technology Industry Review*, (27), 49–76.

Aksnes, D. W., Rørstad, K., Piro, F. N., & Sivertsen, G. (2013). Are mobile researchers more productive and cited than non-mobile researchers? A large-scale study of Norwegian scientists. *Research Evaluation*, *22*(4), 215–223.

Auriol, L. (2006). International mobility of doctorate holders: First results and methodology advances. In *first PRIME Indicators Conference, Lugano*.

Azoulay, P., Ganguli, I., & Zivin, J. S. G. (2016). *The Mobility of Elite Life Scientists: Professional and Personal Determinants* (Working Paper No. 21995). National Bureau of Economic Research. https://doi.org/10.3386/w21995

Baruffaldi, S. H., & Landoni, P. (2012). Return mobility and scientific productivity of researchers working abroad: The role of home country linkages. *Research Policy*, *41*(9), 1655–1665. https://doi.org/10.1016/j.respol.2012.04.005

Børing, P., Flanagan, K., Gagliardi, D., Kaloudis, A., & Karakasidou, A. (2015). International mobility: Findings from a survey of researchers in the EU. *Science and Public Policy*, *42*(6), 811–826. https://doi.org/10.1093/scipol/scv006

Cañibano, C. (2017). Scientific Mobility and Economic Assumptions: From the Allocation of Scientists to the Socioeconomics of Network Transformation. *Science as Culture*, *26*(4), 505–519. https://doi.org/10.1080/09505431.2017.1363173

Cañibano, C., Fox, M. F., & Otamendi, F. J. (2016). Gender and patterns of temporary mobility among researchers. *Science and Public Policy*, *43*(3), 320–331. https://doi.org/10.1093/scipol/scv042

Cañibano, C., Otamendi, F. J., & Solís, F. (2011). International temporary mobility of researchers: a cross-discipline study. *Scientometrics*, *89*(2), 653. https://doi.org/10.1007/s11192-011-0462-2

Cañibano, C., & Woolley, R. (2015). Towards a Socio-Economics of the Brain Drain and Distributed Human Capital. *International Migration*, *53*(1), 115–130. https://doi.org/10.1111/imig.12020

Caron, E., & van Eck, N. J. (2014). Large scale author name disambiguation using rule-based scoring and clustering. In *19th International Conference on Science and Technology Indicators."Context counts: Pathways to master big data and little data"* (pp. 79–86). CWTS-Leiden University Leiden. Retrieved from http://www.researchgate.net/profile/Tindaro_Cicero/publication/265396216_Research_quality_characteristics_of_publications_and_socio-demographic_features_of_Universities_and_Researchers_evidence_from_the_Italian_VQR_2004-2010_evaluation_exercise/links/540d89180cf2df04e754b658.pdf#page=91

Chinchilla-Rodríguez, Z., Miao, L., Murray, D., Robinson-García, N., Costas, R., & Sugimoto, C. R. (2017). Networks of international collaboration and mobility: a comparative study.

Costas, R., van Leeuwen, T. N., & Bordons, M. (2010). A bibliometric classificatory approach for the study and assessment of research performance at the individual level: The effects of age on productivity and impact. *Journal of the American Society for Information Science and Technology*, *61*(8), 1564–1581. https://doi.org/10.1002/asi.21348

Davenport, S. (2004). Panic and panacea: Brain drain and science and technology human capital policy. *Research Policy*, *33*(4), 617–630. https://doi.org/10.1016/j.respol.2004.01.006

Furukawa, T., Shirakawa, N., & Okuwada, K. (2011). Quantitative analysis of collaborative and mobility networks. *Scientometrics*, *87*(3), 451–466. https://doi.org/10.1007/s11192-011-0360-7

Halevi, G., Moed, H. F., & Bar-Ilan, J. (2015). Researchers' Mobility, Productivity and Impact: Case of Top Producing Authors in Seven Disciplines. *Publishing Research Quarterly*, *32*(1), 22–37. https://doi.org/10.1007/s12109-015-9437-0

Halevi, G., Moed, H. F., & Bar-Ilan, J. (2016). Does Research Mobility Have an Effect on Productivity and Impact? *International Higher Education*, *0*(86), 5–6.

Hunter, R. S., Oswald, A. J., & Charlton, B. G. (2009). The Elite Brain Drain*. *The Economic Journal*, *119*(538), F231–F251. https://doi.org/10.1111/j.1468-0297.2009.02274.x

Jonkers, K., & Tijssen, R. (2008). Chinese researchers returning home: Impacts of international mobility on research collaboration and scientific productivity. *Scientometrics*, *77*(2), 309–333. https://doi.org/10.1007/s11192-007-1971-x

Kato, M., & Ando, A. (2017). National ties of international scientific collaboration and researcher mobility found in Nature and Science. *Scientometrics*, *110*(2), 673–694. https://doi.org/10.1007/s11192-016-2183-z

Larivière, V., & Costas, R. (2016). How Many Is Too Many? On the Relationship between Research Productivity and Impact. *PLOS ONE*, *11*(9), e0162709. https://doi.org/10.1371/journal.pone.0162709

Laudel, G. (2003). Studying the brain drain: Can bibliometric methods help? *Scientometrics*, *57*(2), 215–237. https://doi.org/10.1023/A:1024137718393

Levin, S. G., & Stephen, P. E. (1999). Are the Foreign Born a Source of Strength for U . S . Science ? *Science*, *285*(5431), 3–7. https://doi.org/10.1126/science.285.5431.1213

Markova, Y. V., Shmatko, N. A., & Katchanov, Y. L. (2016). Synchronous international scientific mobility in the space of affiliations: evidence from Russia. *SpringerPlus*, *5*, 480. https://doi.org/10.1186/s40064-016-2127-3

Meyer, J.-B. (2001). Network Approach versus Brain Drain: Lessons from the Diaspora. *International Migration*, *39*(5), 91–110. https://doi.org/10.1111/1468-2435.00173

Moed, H. F., Aisati, M., & Plume, A. (2013). Studying scientific migration in Scopus. *Scientometrics*, *94*(3), 929–942. https://doi.org/10.1007/s11192-012-0783-9

Moed, H. F., & Halevi, G. (2014). A bibliometric approach to tracking international scientific migration. *Scientometrics*, 1–15. https://doi.org/10.1007/s11192-014-1307-6

Moguérou, P., di Pietrogiacomo, M. P., Da Costa, O., & Laget, P. (2006). Indicators on researchers' career and mobility in Europe: a "modelling" approach. In *Blue Sky II Concur-rent workshop*.

Nane, G. F., Larivière, V., & Costas, R. (2017). Predicting the age of researchers using bibliometric data. *Journal of Informetrics*, *11*(3), 713–729. https://doi.org/10.1016/j.joi.2017.05.002

OECD. (2008). *The global competition for talent: Mobility of the highly skilled*. Paris: OECD.

OECD. (2010). *Measuring Innovation: A New Perspective*. OECD Publishing.

Robinson-Garcia, N., Cañibano, C., Woolley, R., & Costas, R. (2016). Tracing scientific mobility of Early Career Researchers in Spain and The Netherlands through their publications. *ArXiv:1606.00155 [Cs]*. Retrieved from http://arxiv.org/abs/1606.00155

Scott, P. (2015). Dynamics of academic mobility: Hegemonic internationalisation or fluid globalisation. *European Review*, *23*(S1), S55–S69.

Smalheiser, N. R., & Torvik, V. I. (2009). Author name disambiguation. *Annual Review of Information Science and Technology*, *43*(1), 1–43. https://doi.org/10.1002/aris.2009.1440430113

Stephan, P., Franzoni, C., & Scellato, G. (2016). Global competition for scientific talent: evidence from location decisions of PhDs and postdocs in 16 countries. *Industrial and Corporate Change*, *25*(3), 457–485. https://doi.org/10.1093/icc/dtv037

Sugimoto, C. R., & Lariviere, V. (2018). *Measuring Research: What Everyone Needs to Know®*. Oxford University Press.

Sugimoto, C. R., Robinson-Garcia, N., & Costas, R. (2016). Towards a global scientific brain: Indicators of researcher mobility using co-affiliation data. *ArXiv:1609.06499 [Cs]*. Retrieved from http://arxiv.org/abs/1609.06499

Sugimoto, C. R., Robinson-Garcia, N., Murray, D. S., Yegros-Yegros, A., Costas, R., & Larivière, V. (2017). Scientists have most impact when they're free to move. *Nature*, *550*(7674), 29. https://doi.org/10.1038/550029a

Wagner, C. S., & Jonkers, K. (2017). Open countries have strong science. *Nature*, *550*(7674), 32. https://doi.org/10.1038/550032a